\documentclass[aps,preprint,nofootinbib]{revtex4}
\usepackage{graphicx} 

\newcommand{\nn}{\nonumber}

\newcommand{\GeV}{~\mbox{GeV}}
\newcommand{\MeV}{~\mbox{MeV}}
\newcommand{\KeV}{~\mbox{KeV}}
\newcommand{\eV}{~\mbox{eV}}

\newcommand{\Sp}{\mbox{Sp}}

\newcommand{\br}[1]{\left( #1 \right)}
\newcommand{\brs}[1]{\left[ #1 \right]}
\newcommand{\brf}[1]{\left\{ #1 \right\}}
\newcommand{\brm}[1]{\left| #1 \right|}

\newcommand{\ba}{\begin{eqnarray}}
\newcommand{\ea}{\end{eqnarray}}

\begin{document}

\title{The decay $\phi\to f_0(980) \gamma$ and the
process $e^+e^- \to \phi f_0(980)$}

\author{Yu.~M.~Bystritskiy}
\email{bystr@theor.jinr.ru}
\affiliation{Joint Institute for Nuclear Research, Dubna, Russia}

\author{M.~K.~Volkov}
\email{volkov@theor.jinr.ru}
\affiliation{Joint Institute for Nuclear Research, Dubna, Russia}

\author{E.~A.~Kuraev}
\email{kuraev@theor.jinr.ru}
\affiliation{Joint Institute for Nuclear Research, Dubna, Russia}

\author{E.~Barto\v{s}}
\email{erik.bartos@savba.sk}
\affiliation{Institute of Physics, Slovak Academy of Sciences, Bratislava}

\author{M.~Se\v{c}ansk\'y}
\email{secansky@theor.jinr.ru}
\affiliation{Joint Institute for Nuclear Research, Dubna, Russia}
\affiliation{Institute of Physics, Slovak Academy of Sciences, Bratislava}

\date{\today}

\begin{abstract}
The decay $\phi \to f_0(980) \gamma$ and process $e^+e^-\to \phi
f_0(980)$ are considered within the local Nambu-Jona-Lasinio model.
In the amplitudes of these processes contributions of
$s$-quark and kaon loops are taken into account. The kaon loop
gives a dominant contribution. Our estimation for the decay width of
$\phi \to f_0 \gamma$ is in satisfactory agreement with recent
experimental data. This allows us to make some predictions for
cross sections of the process $e^+e^- \to \gamma^* \to \phi f_0$
which can be tested in the C-$\tau$ factory. The total and
differential cross sections of this process are calculated and
presented in the figures.
\end{abstract}


\maketitle

\section{Introduction}

In the last years a lot of experimental \cite{Achasov:2000ku,
Ambrosino:2005wk} and theoretical
\cite{Achasov:1987ts, Achasov:1998cc, Jaffe:1976ig, Gerasimov:2004kq,
Teshima:2005qr, Fariborz:2007ai, Escribano:2006mb, Volkov:2001ct}
papers have been devoted to the description of $\phi$-meson decays with
the production of scalar isoscalar $f_0$ mesons.

There are different theoretical interpretations of the $f_0(980)$ meson
structure. In papers \cite{Achasov:1987ts, Weinstein:1982gc,Weinstein:1983gd,Weinstein:1990gu},
for example,
this meson is considered as a kaon molecule.
In other papers, this meson is described as a four quark state
\cite{Jaffe:1976ig, Achasov:1987ts} or as an admixture of
quark-antiquark and diquark-antidiquark states
\cite{Gerasimov:2004kq, Teshima:2005qr, Fariborz:2007ai}.
Recently, the decays of $\phi$-mesons were considered within the
ChPT \cite{Escribano:2006mb}.

In this paper, the local Nambu-Jona-Lasinio (NJL) model will be used.
All mesons are treated as quark-antiquark states in this model.
In particular, the $f_0(980)$ meson is the admixture of
light $u\bar u$ and $d \bar d$ and strange $s \bar s$ quarks \cite{Volkov:2001ct}.
In the framework of this model we describe the decay $\phi \to f_0 \gamma$.

This decay channel was successfully described in terms of
one-loop Feynman amplitudes with intermediate state of
$K^+K^-$ mesons \cite{Achasov:1987ts}.

The amplitude of this process in our approach we
express in terms of $s$-quark and kaon loops.
The obtained result is in satisfactory agreement with recent
experimental data \cite{Ambrosino:2006hb}.

Using the same approximations we calculate the total and differential
cross sections for the $e^+e^- \to \gamma^* \to \phi f_0$ process.
A comparison with the results obtained in ChPT approach
\cite{Napsuciale:2007wp} and the recent experimental data
\cite{Aubert:2006bu} are discussed in Conclusion.

\section{Process $\phi \to f_0(980) \gamma$}
\label{SectF0Gamma}

The inner parameters of the NJL model are the
constituent quark masses $m_u = m_d = 263\MeV$, $m_s = 407\MeV$
and the ultraviolet cut-off parameter $\Lambda = 1.27\GeV$
\cite{Volkov:1986zb, Volkov:2001ct}.
These parameters are fixed by a value of the weak pion decay
$\pi\to\mu\nu$ constant $f_\pi = 92.4\MeV$ and
by the strong decay $\rho \to \pi \pi$,
$g_\rho = 5.94$ (that correspond to the width $\Gamma_{\rho \to
\pi \pi} = 149.4\MeV$)
\footnote{Let us note that in \cite{Volkov:1986zb} some other values
of these parameters were used which corresponded to
$f_\pi = 93\MeV$, $g_\rho = 6.14$ (in that case the width
$\Gamma_{\rho \to \pi \pi} = 155\MeV$). Here we use
modern experimental data \cite{Yao:2006px}
for fixing our model parameters.}
\cite{Yao:2006px}.

Besides, we use the angle $\alpha$ that describes the deviation
from the angle of ideal mixing of scalar mesons in
the singlet-octet sector.
In the case of ideal mixing we have two states:
the state $\sigma_u$ consists of light
$u$ and $d$ quarks, and the state $\sigma_s$ consists of
$s$ quarks only.
The angle $\alpha$ allows us to express
physical states $f_0(980)$ and $\sigma$ through
the states $\sigma_u$ and $\sigma_s$:
\ba
    \sigma &=& \sigma_u \cos{\alpha} - \sigma_s\sin{\alpha}, \nn \\
    f_0 &=& \sigma_u \sin{\alpha} + \sigma_s\cos{\alpha}. \nn
\ea
The value of $\alpha=11.85^o$ was obtained by using t'Hooft interaction
and mass difference of $\eta$ and $\eta'$ mesons \cite{Volkov:1999qn, Volkov:1998ax}.

Part of Lagrangian corresponding to a
quark-meson interaction has the form
\ba
    \Delta {\cal L}_{int} =
    \bar q \brf{
    g_{\sigma_u} \lambda_u \sigma_u +
    g_{\sigma_s} \lambda_s \sigma_s +
    i \gamma_5 g_K \br{\lambda_{K^+} K^+ + \lambda_{K^-} K^-} +
    \frac{g_\phi}{2} \gamma_\nu \lambda_s \phi^\nu
    } q.
    \label{QuarkMesonLagrangian}
\ea
Using the parameters of the model it is possible to calculate all
meson-quark coupling constants and the constant corresponding to
additional renormalization of the pseudoscalar fields $Z_\pi$ and
$Z_K$ which takes into account the transition of
pseudoscalar mesons to axial-vector mesons \cite{Volkov:1986zb}
\ba
    &&
    g_{\sigma_u} = \br{4 I\br{m_u,m_u}}^{-1/2} = \frac{g_\rho}{\sqrt{6}} = 2.42,
    \qquad
    g_{\sigma_s} = \br{4 I\br{m_s,m_s}}^{-1/2} = 2.98, \label{ConstantsAndIntegrals} \\
    &&
    g_{K_0^*} = \br{4 I\br{m_u,m_s}}^{-1/2} = 2.71,
    \qquad
    g_{\phi} = \sqrt{6} \, g_{\sigma_s} = 7.32, \nn\\
    &&
    Z_K = \br{1 - \frac{3 {(m_u+m_s)}^2}{2 M_{K_1}^2}}^{-1} = 1.52,
    \qquad
    g_K = g_{K_0^*} Z_K^{1/2} = 3.34, \nn
\ea
where $M_{K_1}=1403\MeV$ is the mass of the strange axial-vector meson $K_1$
and
\ba
    I(m,m) &=& \frac{3}{\br{2\pi}^4}
    \int d^4 k
    \frac{\theta\br{\Lambda^2-k^2}}
    {\br{k^2+m^2}^2} =
    \frac{3}{\br{4\pi}^2}
    \br{
        \ln\br{\frac{\Lambda^2}{m^2}+1}
        -
        \frac{\Lambda^2}{\Lambda^2 + m^2}
    }, \nn \\
    I(m_1,m_2) &=& \frac{3}{\br{2\pi}^4}
    \int d^4 k
    \frac{\theta\br{\Lambda^2-k^2}}
    {\br{k^2+m_1^2}\br{k^2+m_2^2}} = \nn\\
    &=&
    \frac{3}{\br{4\pi}^2 \br{m_2^2-m_1^2}}
    \br{
        m_2^2 \ln\br{\frac{\Lambda^2}{m_2^2}+1}
        -
        m_1^2 \ln\br{\frac{\Lambda^2}{m_1^2}+1}
    }. \nn
\ea
We used in (\ref{QuarkMesonLagrangian})
the following combinations of the Gell-Mann matrices:
\ba
&&
\lambda_u =
\br{
    \begin{array}{ccc}
        1 & 0 & 0 \\
        0 & 1 & 0 \\
        0 & 0 & 0
    \end{array}
}
=\frac{(\lambda_8+\sqrt{2}\lambda_0)}{\sqrt{3}},
\qquad
\lambda_s =
\br{
    \begin{array}{ccc}
      0 & 0 & 0 \\
      0 & 0 & 0 \\
      0 & 0 & -\sqrt{2}
    \end{array}
}
=\frac{(-\lambda_0+\sqrt{2}\lambda_8)}{\sqrt{3}},
\nn\\
&&
\lambda_{K^+} =
\br{
    \begin{array}{ccc}
      0 & 0 & 0 \\
      0 & 0 & 0 \\
      \sqrt{2} & 0 & 0
    \end{array}
}
=\frac{\lambda_4-i\lambda_5}{\sqrt{2}},
\qquad
\lambda_{K^-} =
\br{
    \begin{array}{ccc}
      0 & 0 & \sqrt{2} \\
      0 & 0 & 0 \\
      0 & 0 & 0
    \end{array}
}
=\frac{\lambda_4+i\lambda_5}{\sqrt{2}}.
\nn
\ea

\begin{figure}
\includegraphics[width=1\textwidth]{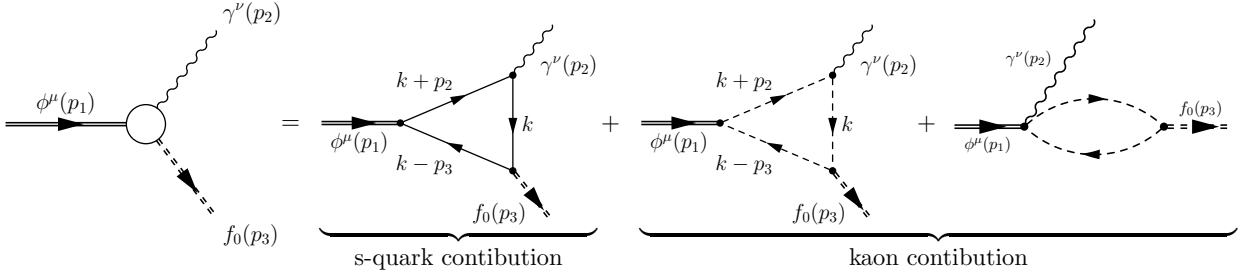}
\caption{Feynman diagrams describing the amplitude of the decay $\phi \to f_0 \gamma$.}
\label{Fig:1}
\end{figure}

The decay $\phi \to f_0 \gamma$ in the local NJL model
is described by the diagrams shown in Fig. \ref{Fig:1},
where the first diagram contains the $s$-quark loop
and presents the contribution of the first order of $1/N_c$ expansion
(where $N_c=3$ is the number of quark colors), and the last two diagrams
describes the contribution of kaon loops (next order of $1/N_c$ expansion).
All vertices of these diagrams were calculated in terms of the quark loops.
Only the divergent part of this quark loop integrals
with the appropriate ultraviolet regularization with the cut-off
parameter $\Lambda$ was taken into account.
As a result, all coupling constants in formula (\ref{ConstantsAndIntegrals})
was calculated.
Let us consider, for example, the calculation of the vertex $f_0 K^+ K^-$.
The integral corresponding to this vertex has the form
\ba
    &&
    g_{f_0 K^+ K^-} =
    i \frac{3}{\br{2\pi}^4}
    g_K^2 \int dk \times \nn\\
    &&\quad \times
    \left\{
        \Sp\brs{\br{\lambda_s g_{\sigma_s} \cos \alpha + \lambda_u g_{\sigma_u} \sin \alpha}
         S\br{k-p_+} \lambda_{K^+} \gamma_5 S\br{k} \lambda_{K^-} \gamma_5 S\br{k+p_-}}_f
    \right.
    +\nn\\
    &&\qquad+\left.
        \Sp\brs{\br{\lambda_s g_{\sigma_s} \cos \alpha + \lambda_u g_{\sigma_u} \sin \alpha}
         S\br{k-p_-} \lambda_{K^-} \gamma_5 S\br{k} \lambda_{K^+} \gamma_5 S\br{k+p_+}}_f,
    \right\}
\ea
where $S\br{k}$ is the matrix of quark propagators:
\ba
    S\br{k} = \mbox{diag}\br{\frac{\hat k + m_u}{k^2 - m_u^2}, \frac{\hat k + m_d}{k^2 - m_d^2}, \frac{\hat k + m_s}{k^2 - m_s^2}},
\ea
and $\Sp[\dots]_f$ is the trace over flavour indices. Calculation of this trace leads to
the following expression:
\ba
    g_{f_0 K^+ K^-} &=&
   i \frac{3}{\br{2\pi}^4}
    g_K^2
    \int dk \times\nn\\
        &\times&
        \left\{
            \br{-2\sqrt{2}} g_{\sigma_s} \cos\alpha
            \frac{
                    \Sp\brs{\br{\hat k-\hat p_+ + m_s} \gamma_5 \br{\hat k + m_u} \gamma_5 \br{\hat k+\hat p_-+m_s}}
                }
                {
                    \br{\br{k- p_+}^2 - m_s^2} \br{k^2 - m_u^2} \br{\br{k+ p_-}^2-m_s^2}
                }
        \right. +\nn\\
        &&\quad+
        \left.
            2 g_{\sigma_u} \sin\alpha
            \frac{
                    \Sp\brs{\br{\hat k-\hat p_+ + m_u} \gamma_5 \br{\hat k + m_s} \gamma_5 \br{\hat k+\hat p_-+m_u}}
                }
                {
                    \br{\br{k- p_+}^2 - m_u^2} \br{k^2 - m_s^2} \br{\br{k+ p_-}^2-m_u^2}
                }
        \right\}.
        \nn
\ea
After calculation of the trace over the Dirac matrices we separate out
the divergent terms of the quark loop and calculate them in Euclidean metric:
\ba
    g_{f_0 K^+ K^-}
        &=&
        i \frac{3}{\br{2\pi}^4}
        g_K^2
        \int \frac{dk}{i \pi^2}
        \left\{
            \br{-2\sqrt{2}} g_{\sigma_s} \cos\alpha
            \frac{
                    4\br{ \br{k^2-m_u^2} \br{m_u - 2 m_s} + \br{\mbox{finite terms}} }
                }
                {
                    \br{k^2 - m_u^2} \br{k^2-m_s^2}^2
                }
        \right. +\nn\\
        &&\qquad\qquad\qquad\qquad +
        \left.
            2 g_{\sigma_u} \sin\alpha
            \frac{
                    4\br{ \br{k^2-m_u^2} \br{m_s - 2 m_u} + \br{\mbox{finite terms}} }
                }
                {
                    \br{k^2 - m_s^2} \br{k^2-m_u^2}^2
                }
        \right\} =\nn\\
        &=&
        g_K^2
        \left\{
            \br{-2\sqrt{2}} g_{\sigma_s} \cos\alpha \br{m_u - 2 m_s}
            \br{4 \frac{3}{\br{2\pi}^4} \int
                \frac{dk}{ \br{k^2 + m_s^2} \br{k^2+m_s^2} } }_1 +
        \right. \nn \\
        &&\qquad\qquad\quad+ \left.
            2 g_{\sigma_u} \sin\alpha \br{m_s - 2 m_u}
            \br{4 \frac{3}{\br{2\pi}^4} \int
                \frac{dk}{ \br{k^2 + m_u^2} \br{k^2+m_s^2} } }_2
        \right\}.
\ea
Recalling (\ref{ConstantsAndIntegrals}) the expression in the
round brackets can be rewritten as
$(...)_1=4 I\br{m_s,m_s}=\br{g_{\sigma_s}}^{-2}$
and
$(...)_2=4 I\br{m_u,m_s}=\br{g_{K_0^*}}^{-2}$
and the vertex obtains the form
\ba
    g_{f_0 K^+ K^-}
        &=&
        2
        \brf{
            \sqrt{2} g_{\sigma_s} \cos\alpha \br{2 m_s-m_u}
            \br{\frac{g_K}{g_{\sigma_s}}}^2
            -
            g_{\sigma_u} \sin\alpha \br{2 m_u - m_s}
            \br{\frac{g_K}{g_{K_0^*}}}^2
            } = \nn\\
        &=& 5.51\GeV.
\ea
Similar calculations give us the following expressions for other constants:
\ba
g_{\phi^\mu K^+ K^-} &=& \frac{g_\phi}{\sqrt{2}} \br{\frac{g_{K_0^*}}{g_{\sigma_s}}}^2 Z_K \br{p^+ + p^-}^\mu,\nn\\
g_{A^\mu K^+ K^-} &=& e \br{p^+ + p^-}^\mu,
\ea
where $p^\pm$ are the $K^\pm$ momenta and $e$ is the electric charge ($e^2/4\pi = 1/137$).
Let us note that in the vertices $g_{\phi^\mu K^+ K^-}$ and
$g_{\phi^\mu \gamma\nu K^+ K^-}$ the factor $Z_K$ disappears after taking
into account $K^+ \to K^+_1$ transitions on the kaon line. A similar situation
takes place in the decay of $\rho \to \pi \pi$ \cite{Volkov:1986zb}.
The vertex $g_{\phi^\mu K^+ K^-}$ was derived in \cite{Volkov:1993jw, Ebert:1994mf}
and leads to satisfactory
agreement with the experiment -- we get the decay width
${\Gamma_{\phi \to K K}=1.88}\MeV$ while the experimental value is
$\Gamma^{exp}_{\phi \to K K}=2.1\MeV$ \cite{Yao:2006px}.

Now we can calculate the contributions of the quark and kaon loops
(see Fig. \ref{Fig:1}) to the process
$\phi \to f_0 \gamma$. The quark loop gives the amplitude
\ba
M^{(s)}_{\phi \to f_0 \gamma} &=&
C^{(s)} A^{(s)}_{\phi \to f_0 \gamma}
\br{g^{\mu\nu} \br{p_1 p_2} - p_1^\nu p_2^\mu} e_\mu(p_1) e_\nu(p_2), \label{QuarkAmplitude} \\
C^{(s)} &=& \frac{e}{(4 \pi)^2} g_\rho g_{\sigma_s} \cos{\alpha}, \nn\\
A^{(s)}_{\phi \to f_0 \gamma} &=&
\int\limits_0^1 dx \int\limits_0^{1-x} dy \frac{8 m_s \br{4 x y-1}}{m_s^2 - y(1-y) M_\phi^2 + x y \br{M_\phi^2-M_{f_0}^2} + i\epsilon}.
\nn
\ea
Following the quark confinement condition we take into account
only the real part of this amplitude. Then the amplitude square is
\ba
    \brm{Re\br{M^{(s)}_{\phi \to f_0 \gamma}}}^2 =
    \frac{1}{2} \br{M_\phi^2 - M_{f_0}^2}^2
    \brm{ C^{(s)} Re\br{A^{(s)}_{\phi \to f_0 \gamma}}}^2.
\ea
It gives the following contribution to the decay width:
\ba
    \Gamma^{(s)}_{\phi\to f_0 \gamma} &=&
    \frac{1}{2^5 3 \pi}
    \frac{\br{M_\phi^2 - M_{f_0}^2}^3}{M_\phi^3}
    \brm{ C^{(s)} Re\br{A^{(s)}_{\phi \to f_0 \gamma}}}^2 = 6.75 \eV.
    \label{QuarkContibution}
\ea
%
The kaon loop gives the amplitude
\ba
M^{(K)}_{\phi \to f_0 \gamma} &=&
C^{(K)} A^{(K)}_{\phi \to f_0 \gamma}
\br{g^{\mu\nu} \br{p_1 p_2} - p_1^\nu p_2^\mu} e_\mu(p_1) e_\nu(p_2), \label{KaonAmplitude} \\
C^{(K)} &=& \frac{e}{(4 \pi)^2} \frac{g_\rho}{\sqrt{2}} g_{f_0 K^+K^-}, \nn\\
A^{(K)}_{\phi \to f_0 \gamma} &=&
\int\limits_0^1 dx \int\limits_0^{1-x} dy \frac{8 \br{4 x y}}{M_K^2 - y(1-y) M_\phi^2 + x y \br{M_\phi^2-M_{f_0}^2} + i\epsilon}.
\nn
\ea
Its contribution to the decay width $\phi \to f_0 \gamma$
is dominant
\ba
\Gamma^{(K)}_{\phi\to f_0 \gamma} &=& \frac{1}{2^5 3 \pi}
\frac{\br{M_\phi^2 - M_{f_0}^2}^3}{M_\phi^3}
\brm{C^{(K)} A^{(K)}_{\phi \to f_0 \gamma}}^2.
\ea
It is worth noticing that a theoretical prediction has a strong dependence on
the mass of the $f_0$-meson value. Experimental value is $M_{f_0} = 980 \pm 10\MeV$
and changing $M_{f_0}$ in the interval $970\MeV \leq M_{f_0} \leq 990\MeV$
we obtain the following interval for a theoretical prediction of
decay width
$2.39\KeV \geq \Gamma^{(K)}_{\phi\to f_0 \gamma} \geq 0.66 \KeV $.
With the contribution of the quark loop taken into account
these values slightly change
\ba
\Gamma^{(K+s)}_{\phi\to f_0 \gamma} \approx 0.65\KeV.
\ea
The experimental value is $\Gamma^{(exp)}_{\phi\to f_0 \gamma}= 0.47 \pm 0.03\KeV$ \cite{Ambrosino:2006hb}.
So we can see that our prediction is in qualitative
agreement with experiment at $M_{f_0}=990\MeV$.

\section{Subprocess $\gamma^* \to \phi f_0(980)$}

Let us now consider the cross process for the $\phi\to f_0 \gamma$ decay,
namely, the $\gamma^* \to \phi f_0$. Due to the off-mass-shell photon
here the additional gauge invariant Lorentz structure appears and
the amplitude can be written in the form
\ba
&&M\br{\gamma^*(p_2,\nu) \to \phi(p_1,\mu) f_0 (p_3)}= \nn\\
&&\qquad\qquad=
\sum\limits_{i=s,K}
\frac{C_{(i)}}{2^4 \pi^2}
e_\mu(p_1) e_\nu(p_2)
\br{
    A_{(i)} R_{(1)}^{\mu\nu} +
    B_{(i)} R_{(2)}^{\mu\nu}
},
\label{SubprocessAmplitude}
\ea
where $i$ denotes the type of a contribution ($i=s$ corresponds to
the $s$-quark loop contribution and $i=K$ to the kaon loop
contribution). Two gauge invariant structures
$R_{(1,2)}^{\mu\nu}$ are
\ba
R_{(1)}^{\mu\nu} &=& g^{\mu\nu} \br{p_1 p_2} - p_1^\nu p_2^\mu, \nn\\
R_{(2)}^{\mu\nu} &=&
\br{p_1 - p_2 \frac{p_1^2}{(p_1 p_2)}}^\mu
\br{p_2 - p_1 \frac{p_2^2}{(p_1 p_2)}}^\nu, \nn\\
&&
p_1^\mu R^{(i)}_{\mu\nu} =
p_2^\nu R^{(i)}_{\mu\nu} = 0, \qquad i=1,2.
\ea
The quantities $A_{(i)}$, $B_{(i)}$  in (\ref{SubprocessAmplitude})
depend only on momentum squares ($p_1^2$, $p_2^2$, $p_3^2$)
and $C_{(i)}$ are the product of the coupling constants.

Quark-loop contribution takes the form (which differs from
the case of $\phi\to f_0 \gamma$ by nonzero virtuality of photon)
\ba
C_{(q)} &=& e \, g_\rho \, g_{\sigma_s} \cos \alpha, \nn\\
A_{(q)} &=& -\alpha_{(q)} + \beta_{(q)} \frac{p_1^2 p_2^2}{\br{p_1 p_2}^2}, \nn \\
B_{(q)} &=& \beta_{(q)}, \nn \\
\alpha_{(q)} &=&
\int\limits_0^1 dx \int\limits_0^{1-x} dy
\frac{8 m_s \br{4 x y - 1}}{m_s^2 - x \, z \, p_1^2 - y \, z \, p_2^2 - x \, y \, p_3^2 - i\epsilon}, \nn\\
\beta_{(q)} &=&
\int\limits_0^1 dx \int\limits_0^{1-x} dy
\frac{8 m_s \br{2(x+y) - 4 x y - 1}}{m_s^2 - x \, z \, p_1^2 - y \, z \, p_2^2 - x \, y \, p_3^2 - i\epsilon}, \nn
\ea
where $z = 1-x-y$.
The kaon-loop contribution reads as:
\ba
C_{(K)} &=& e \, g_{\phi K^+ K^-} \, g_{f_0 K^+ K^-}, \nn\\
A_{(K)} &=& -\alpha_{(K)} + \beta_{(K)} \frac{p_1^2 p_2^2}{\br{p_1 p_2}^2}, \nn \\
B_{(K)} &=& \beta_{(K)}, \nn \\
\alpha_{(K)} &=&
\int\limits_0^1 dx \int\limits_0^{1-x} dy
\frac{8 \br{4 x y}}{M_K^2 - x \, z \, p_1^2 - y \, z \, p_2^2 - x \, y \, p_3^2 - i\epsilon}, \nn\\
\beta_{(K)} &=&
\int\limits_0^1 dx \int\limits_0^{1-x} dy
\frac{8 \br{2(x+y) - 4 x y - 1}}{M_K^2 - x \, z \, p_1^2 - y \, z \, p_2^2 - x \, y \, p_3^2 - i\epsilon}. \nn
\ea

\section{Process $e^+ e^- \to \gamma^* \to \phi f_0(980)$}

\begin{figure}
\includegraphics[width=0.5\textwidth]{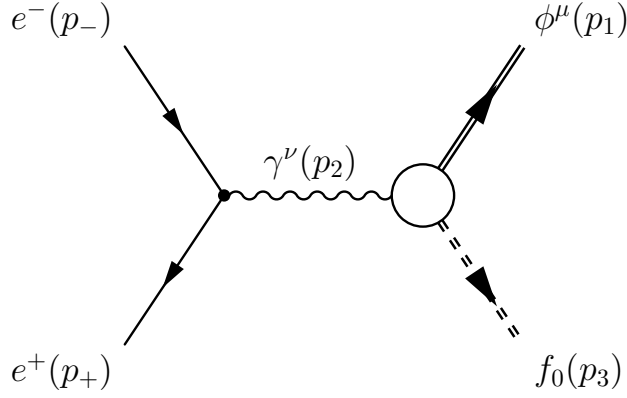}
\caption{Feynman diagram of the process $e^+ e^- \to \gamma^* \to \phi f_0(980)$.}
\label{Fig:2}
\end{figure}

Using the amplitude (\ref{SubprocessAmplitude})
we can write the amplitude for the process $e^+ e^- \to \gamma^* \to \phi f_0(980)$
(see Fig.~\ref{Fig:2})
\ba
&&
M\br{e^+(p_+) e^-(p_-) \to \gamma^*(p_2) \to \phi(p_1) f_0(p_3)}= \nn\\
&& \qquad\qquad
=\frac{4\pi\alpha}{s} \,
J_\mu^{QED} e_\nu(p_2)
\sum\limits_{i=s,K}
\frac{C_{(i)}}{2^4 \pi^2}
\br{
    A_{(i)} R_{(1)}^{\mu\nu} +
    B_{(i)} R_{(2)}^{\mu\nu}
},
\label{eeAmplitude}
\ea
where $J_\mu^{QED}=\bar{v}(p_+)\gamma_\mu u(p_-)$ is
the electromagnetic current of electron and positron annihilation
($J_{\mu} p_2^\mu=0$), and $e_\nu(p_1)$ is the polarization
4-vector of the $\phi-$meson ($p_1^\nu e_\nu(p_1)=0$).

The square modulus of the amplitude (\ref{eeAmplitude})
after summation over polarization states has the form
\ba
\sum_{pol} |M|^2 = \frac{8\pi\alpha}{s}
\brf{\frac{s_1^2}{4} |A|^2 -
\frac{1}{2}\br{|A-\tilde B|^2 s - |\tilde B|^2 \frac{s_1^2}{4M_{\phi}^2}}
\br{E_\phi^2 \br{1-\beta_\phi^2 c^2} - M_{\phi}^2}},
\label{SquareOfAmplitude}
\ea
where
$s_1 = 2(p_1 p_2)=s+M_\phi^2-M_{f_0}^2$,
$\tilde B = B (4 s M_\phi^2/s_1^2)$,
$E_\phi = \br{s+M^2_\phi-M_{f_0}^2}/\br{2\sqrt{s}}$
is the $\phi$-meson energy in the center-of-mass system
$c=\cos{\theta}=\cos(\vec{p}_-,\vec{p}_1)$ is
the cosine of the emission angle of the $\phi$-meson, and
$\beta_\phi=\sqrt{\lambda\br{s,M_\phi^2,M_{f_0}^2}}/\br{s+ M_\phi^2- M_{f_0}^2}$
is the velocity of the $\phi$-meson
($\lambda\br{x,y,z} = x^2 + y^2 + z^2 - 2 x y - 2 x z - 2 y z$ is the well-known
triangle function). The quantities $A$ and $B$ in (\ref{SquareOfAmplitude})
are the sums of quark and kaon contributions
\ba
A &=& C_{(q)} A_{(q)} + C_{(K)} A_{(K)}, \nn\\
B &=& C_{(q)} B_{(q)} + C_{(K)} B_{(K)}. \nn
\ea
The phase volume is
\ba
d \Gamma_2 = \frac{d^3 p_1}{2 E_\phi} \, \frac{d^3 p_3}{2 E_{f_0}}
\, \frac{(2\pi)^4}{(2\pi)^6} \, \delta^4\br{p_+ + p_- - p_1 - p_3} =
\frac{\sqrt{ \lambda\br{s,M_\phi^2,M_{f_0}^2} }}{16\pi s} \, dc.
\ea
The differential cross section can be written in the form
\ba
\frac{d\sigma^{e^+ e^- \to \phi f_0}}{d\cos\theta} =
\frac{\pi\alpha^2}{s} \br{D(s)+E(s)\cos^2\theta},
\label{CrossSection}
\ea
where
\ba
    D(s) &=& \frac{4\pi \sqrt{ \lambda\br{s,M_\phi^2,M_{f_0}^2} }}
                {2^7 \pi s^2 \alpha}
                \brf{
                    s_1^2 |A|^2 -
                    2 \br{|A-\tilde B|^2 s - |\tilde B|^2 \frac{s_1^2}{4M_{\phi}^2}}
                    \br{E_\phi^2 - M_{\phi}^2}
                },
                \label{Dfunc}\\
    E(s) &=& \frac{4\pi \sqrt{ \lambda\br{s,M_\phi^2,M_{f_0}^2} }}
                {2^7 \pi s^2 \alpha}
                2 \beta_\phi^2 E_\phi^2
                \br{|A-\tilde B|^2 s - |\tilde B|^2 \frac{s_1^2}{4M_{\phi}^2}}
                .
                \label{Efunc}
\ea
The total cross section then reads as:
\ba
    \sigma(s) =
    \frac{2 \alpha^2}{s}
    \br{D(s) + \frac{1}{3} E(s)}.
\label{TotalCrossSection}
\ea

\begin{figure}
\includegraphics[width=0.8\textwidth]{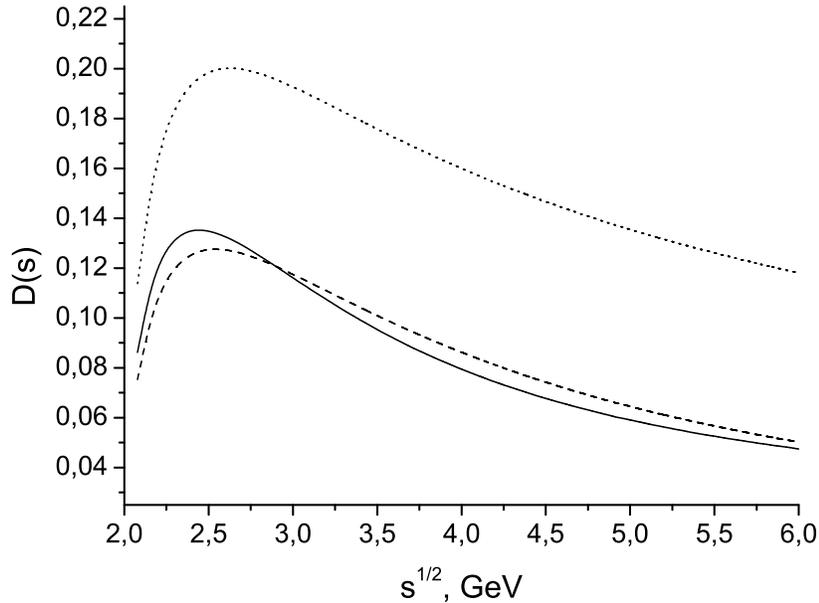}
\caption{Coefficient function $D(s)$ from
differential cross section (see (\ref{Dfunc})).
The dotted line is the quark loop contribution, dashed line
is the kaon loop contribution and the solid line is
the total value of $D(s)$.}
\label{DPlot}
\end{figure}

\begin{figure}
\includegraphics[width=0.8\textwidth]{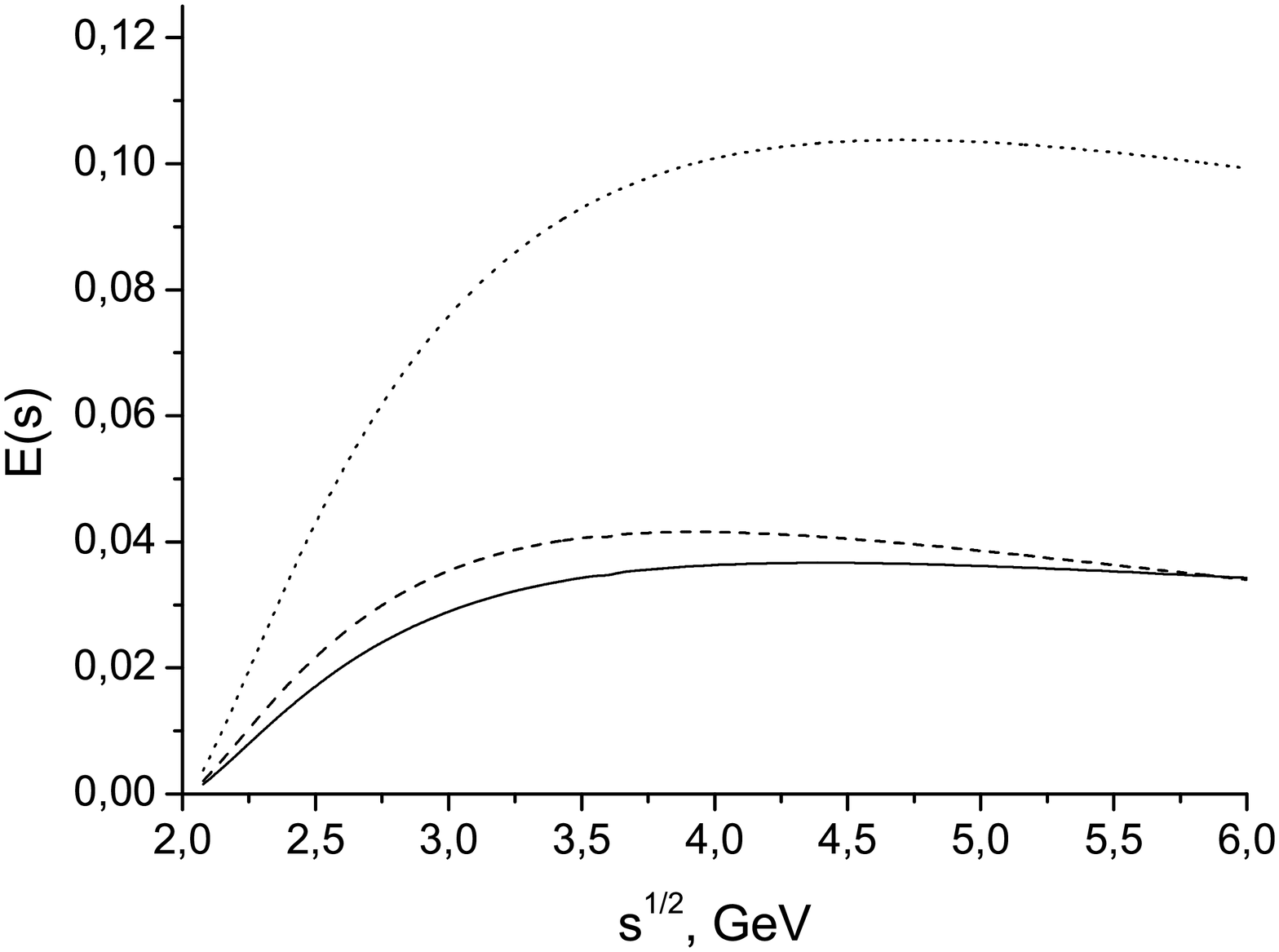}
\caption{Coefficient function $E(s)$ from
differential cross section (see (\ref{Efunc})).
The dotted line is the quark loop contribution, dashed line
is the kaon loop contribution and the solid line is
the total value of $E(s)$.}
\label{EPlot}
\end{figure}

\begin{figure}
\includegraphics[width=0.8\textwidth]{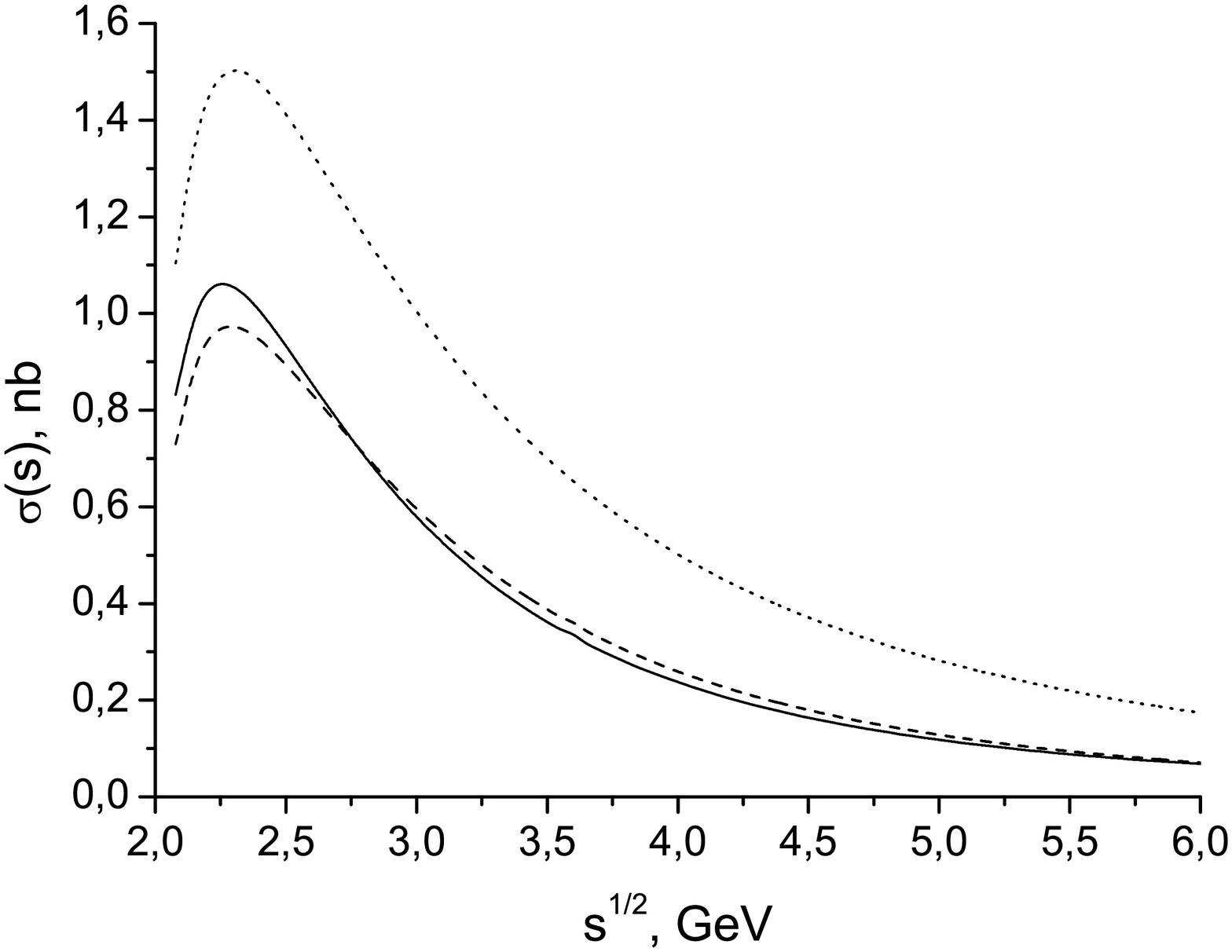}
\caption{Total cross section of the $e^+ e^- \to \gamma^* \to \phi f_0(980)$
process (see (\ref{TotalCrossSection})).
The dotted line is the quark loop contribution, dashed line
is the kaon loop contribution and the solid line is
the total value of cross section.}
\label{TotalPlot}
\end{figure}

Unlike the $\phi \to f_0 \gamma$ decay, where
the contribution of the quark loop was negligible, in the process
$e^+ e^- \to \gamma^* \to \phi f_0(980)$ both contributions
are of the same order.
In Figs. \ref{DPlot} and \ref{EPlot} we show the contributions
of quarks and kaons separately for the
values of $D(s)$ and $E(s)$.
In Fig. \ref{TotalPlot}, the same contributions are shown for the
total cross section.

\section{Conclusion}

The decay $\phi \to f_0(980) \gamma$ was calculated in the framework of
local NJL model.
We suppose that all mesons are the quark-antiquark states.
It turns out that the lowest order of $1/N_c$ expansion (where
$N_c=3$ is the number of colors, Hartree-Fock approximation) where only
quark loops are taken into account does not give
satisfactory agreement with the experimental data (see (\ref{QuarkContibution})).
In the next order of $1/N_c$ expansion we have to consider the meson loops and
they give the dominant contribution to the amplitude of the process
$\phi \to f_0 \gamma$, which leads to satisfactory agreement
with experimental data. By the way, a similar approximation
was also used in other models for description of this process
(see \cite{Achasov:1987ts, Weinstein:1990gu,Escribano:2006mb}).

In the same approximation of the local NJL model the total probability and
the differential cross section of the process $e^+e^- \to \phi f_0(980)$ were calculated.

The recent experiment \cite{Aubert:2006bu} of production
$K^+K^- \pi^+\pi^-$ in annihilation
channel at high energy of $e^+e^-$ collision
show some structure of $0.7~\mbox{nb}$ size  in the
region $\sqrt{s}\approx 2.175\GeV$, which was treated as a resonance state.
The value of the cross section exceeds the theoretical
(non-resonant) cross section calculated in frames of ChPT
which is equal $0.15 nb$. Our result exceeds the
experimental value by a factor 1.2 in this energy range.
The difference can be associated with the background from the
channel $e^+e^-\to \phi \pi^+\pi^-$ with the
effective mass of $\pi^+\pi^-$ outside the $f_0$ meson width.

The investigation of this process and the set of similar ones
with production of heavy and radially excited mesons could be part of the physical program
of the BABAR and the BES-III experiment.

\newpage
\begin{acknowledgments}
One of us (E.K.) is grateful to the Institute of Physics, SAS for
hospitality and
Grant INTAS 05-1000008-8528. The
work was partly supported also by the Slovak Grant Agency for
Sciences VEGA, Grant No. 2/7116/27. M.K.V. acknowledges
the support of grant RFBR (no. 05-02-16699).
\end{acknowledgments}


\end{document}